\documentclass[12pt]{article}

\usepackage{scicite}
\usepackage{times}
\usepackage{graphicx}
\usepackage{pdfpages}

\def\ale{\mathrel{\hbox{\rlap{\hbox{\lower4pt\hbox{$\sim$}}}\hbox{$<$}}}}
\def\age{\mathrel{\hbox{\rlap{\hbox{\lower4pt\hbox{$\sim$}}}\hbox{$>$}}}}

\usepackage{color}
\def\simless{\ale}
\def\lesssim{\ale}
\def\gtrsim{\age}
\newenvironment{sciabstract}{%
\begin{quote} \bf}
{\end{quote}}

\newcommand{\ubslong}{Swift J164449.3+573451}
\newcommand{\ubs}{Sw 1644+57}

\newcommand{\z}{0.3543}

\newcommand{\sun}{\odot}

\def\apjl{ApJL }
\def\aj{AJ }
\def\apj{ApJ }
\def\apjl{ApJL }

\def\araa{ARA\&A }
\def\aap{A\&A }
\def\nat{Nature }

\def\mnras{MNRAS}
\def\apss{A\&SS }

\newcounter{lastnote}
\newenvironment{scilastnote}{%
\setcounter{lastnote}{\value{enumiv}}%
\addtocounter{lastnote}{+1}%
\begin{list}%
{\arabic{lastnote}.}
{\setlength{\leftmargin}{.22in}}
{\setlength{\labelsep}{.5em}}}
{\end{list}}

\def\ucb{1}
\def\princeton{2}
\def\einstein{3}
\def\ucl{4}
\def\warwick{5}
\def\tac{6}
\def\ucsc{7}
\def\unam{8}
\def\lbl{9}
\def\ral{10}
\def\stsci{11}
\def\nasa{12}

\title{A relativistic jetted outburst from a massive black hole fed by a 
tidally disrupted star} 

\author{
Joshua S. Bloom$^{\ucb,\ast}$,
Dimitrios Giannios$^{\princeton}$, Brian D. Metzger$^{\princeton,\einstein}$, \\
S. Bradley Cenko$^{\ucb}$, Daniel A. Perley$^{\ucb}$, Nathaniel R. Butler$^{\ucb,\einstein}$, \\ Nial R. Tanvir$^{\ucl}$, Andrew J. Levan$^{\warwick}$, Paul T. O' Brien$^{\ucl}$, \\
Linda E. Strubbe$^{\ucb,\tac}$, Fabio De Colle$^{\ucsc}$, Enrico Ramirez-Ruiz$^{\ucsc}$, \\
William H. Lee$^{\unam}$, Sergei Nayakshin$^{\ucl}$, Eliot Quataert$^{\ucb,\tac}$, \\ 
Andrew R. King$^{\ucl}$, Antonino Cucchiara$^{\ucb,\lbl}$,  James Guillochon$^{\ucsc}$, \\
Geoffrey C. Bower$^{\ral,\ucb}$, Andrew S. Fruchter$^{\stsci}$, Adam N. Morgan$^{\ucb}$, \\
Alexander J. van der Horst$^{\nasa}$, \\
\small{$^{\ucb}$Department of Astronomy, University of California, Berkeley, CA 
94720-3411, USA}\\
\small{$^{\princeton}$Department of Astrophysical Sciences, Peyton Hall, 
Princeton University, Princeton, NJ 08544, USA}\\
\small{$^{\einstein}$NASA Einstein Fellow}\\
\small{$^{\ucl}$Department of Physics and Astronomy, University of Leicester, 
University Road, Leicester LE1 7RH, UK}\\
\small{$^{\warwick}$Department of Physics, University of Warwick, Coventry CV4 
7AL, UK}\\
\small{$^{\tac}$Theoretical Astrophysics Center, University of California, 
Berkeley, CA 94720, USA}\\
\small{$^{\ucsc}$Astronomy and Astrophysics Department, University of 
California, Santa Cruz, CA 95064, USA}\\
\scriptsize{$^{\unam}$Instituto de Astronom\'{\i}a, Universidad Nacional Autonoma de M\'{e}xico, Apdo.~Postal 70--264,
Cd.~Universitaria, M\'{e}xico DF 04510}\\
\scriptsize{$^{\lbl}$Computational Cosmology Center, Lawrence Berkeley National 
Laboratory, 1 Cyclotron Road, Berkeley, CA 94720, USA}\\
\scriptsize{$^{\ral}$Radio Astronomy Laboratory, University of California, Berkeley, 
601 Campbell Hall 3411, Berkeley, CA 94720, USA}\\
\small{$^{\stsci}$Space Telescope Science Institute, 3700 San Martin Drive, 
Baltimore, MD 21218, USA} \\
\small{$^{\nasa}$Universities Space Research Association, NSSTC, 320 Sparkman Drive, Huntsville, AL 35805, USA} \\
\small{$^\ast$To whom correspondence should be addressed; E-mail:  jbloom@astro.berkeley.edu.}
}

\date{}

\begin{document} 
\baselineskip24pt
\maketitle 
\begin{sciabstract}
While gas accretion onto some massive black holes (MBHs) at the centers of galaxies actively powers luminous emission, the vast majority of MBHs are considered dormant. Occasionally, a star passing too near a MBH is torn apart by gravitational forces, leading to a bright panchromatic tidal disruption flare (TDF) \cite{1988Natur.333..523R,1989ApJ...346L..13E,2004ApJ...600..149W,2009MNRAS.400.2070S}. While the high-energy transient \ubslong\ (``\ubs'') initially displayed none of the theoretically anticipated (nor previously observed) TDF characteristics, we show that the observations \cite{levan2011} suggest a sudden accretion event onto a central MBH of mass $\sim10^6-10^7$ solar masses. We find evidence for a mildly relativistic outflow, jet collimation, and a spectrum characterized by synchrotron and inverse Compton processes; this leads to a natural analogy of \ubs\ with a smaller-scale blazar. The phenomenologically novel \ubs\ thus connects the study of TDFs and active galaxies, opening a new vista on disk-jet interactions in BHs and magnetic field generation and transport in accretion systems.
\end{sciabstract}


While variability is common to all active galactic nuclei (AGN)---fundamentally tied to the unsteady accretion flow of gas towards the central MBH---the timescale for active MBHs to dramatically change accretion rates (leading the source to, for example, turn ``off''), is much longer than a human lifetime. The most variable AGN are a subclass called blazars, with typical masses $M_{\rm BH} \approx 10^8-10^9 M_\odot$ ($M_\odot$ is the mass of the Sun), originally found to be radio and optically bright but with luminosities dominated by X rays and gamma rays. Significant changes in the apparent luminosity over minutes- to hour-long timescales are thought to be predominately due to Doppler-beamed emitting regions within a jetted outflow moving relativistically toward the observer \cite{1997ARA&A..35..445U,2003ApJ...593..667M}. The typical Lorentz factor inferred is $\Gamma_{\rm j} = (1 - \beta^2)^{-1/2} \approx 10$, with velocity $v = \beta c$ of the jetted outflow. The high-energy emission is thought to be due to inverse Compton upscattering of the accretion disk photons, photons from within the jet itself, and/or photons from structures external to the accretion disk \cite{2007Ap&SS.309...95B,Fossati+98}. 

Inactive MBHs can suddenly ``turn on'' while being fed by temporary mass accretion established following the tidal disruption of a passing star \cite{1988Natur.333..523R}. If a star of mass $M_*$ and radius $R_*$ passes within the disruption radius $r_d \approx R_* (M_{\rm BH}/M_*)^{1/3} \approx 5 M_7^{-2/3} r_s$ (with $M_{\rm BH} = 10^7 M_7 M_\odot $ and $r_s=2GM_{\rm BH}/c^2$ the
Schwarzschild radius of the BH, $M_*=M_\odot$, $R_*=R_{\odot}$), then a mass of $\sim M_*/2$ will accrete onto the MBH with a peak accretion rate on a timescale of weeks \cite{1989ApJ...346L..13E}. Importantly, the accretion rate for typical scenarios with a $M_7$ BH can be super-Eddington for months \cite{2009ApJ...697L..77R,2009MNRAS.400.2070S}. Candidate TDFs have been observed at X-ray, ultraviolet, and optical wavebands \cite{1999A&A...349L..45K,2008ApJ...676..944G,2010arXiv1009.1627V,2011arXiv1103.0779C}, but the rates are very low [$\sim10^{-5}$ yr gal$^{-1}$; \cite{2004ApJ...600..149W}] and the connection between the observed light curves and spectra to theoretical expectations has been tenuous.

Recently, Giannios \& Metzger \cite{2011arXiv1102.1429G} suggested that a $\sim1$ yr radio event could follow a TDF arising from a jetted relativistic outflow as it interacted with (and was slowed down by) the external ambient medium, akin to the afterglow from external shocks following gamma-ray bursts \cite{rm92}. The supposition was that the observer viewed the event off-axis from the relativistic jet. Just what would be seen if instead the jet were pointed nearly towards the observer---as in the geometry inferred for blazars---was not considered.

\section*{\ubs: a relativistic outflow generated by an accreting massive black hole}

\ubs\ was initially discovered as a long-duration gamma-ray burst (GRB\,110328A) by the Swift satellite \cite{Cummings2011} at a time $t_0$ = 2011 March 28 12:57:45 UT but, given the longevity and flaring of the X-ray afterglow, it was quickly realized that the high-energy emission was unlike that associated with any GRB previously; in a companion paper, Levan et al.\ \cite{levan2011} (who find redshift of the host galaxy of \ubs\ to be $z=$\z) describe in detail the differences between this event and the GRB population observed by Swift. Based on the data available in the first two days following the event, we \cite{bloom2011} first suggested\footnote{A similarly detailed sketch was also posted \cite{2011arXiv1104.2528B} later.} a possible analogy of \ubs\ with a scaled-down version of a blazar impulsively fed by $\sim1 M_\odot$ of stellar mass. 

There are several lines of evidence to suggest an accreting MBH origin. First, the astrometric coincidence of the X-ray, optical, infrared, and radio transient with the light-centroid of the putative host galaxy is strongly indicative of a positional connection to an MBH\footnote{Indeed we predicted the precise astrometric coincidence in \cite{bloom2011} before the connection was solidified \cite{berger2011}. However, within the uncertainties from Hubble imaging [$\sim$300 pc; \cite{levan2011}], the central stellar and gas density could be high enough to allow other progenitors, such as supernovae.}. Second, the compact nature of the associated radio source suggests an emission region no larger than 5 parsec which disfavors other possible extended emission sources (see SOM).  Last, the observed correlation between the X-ray flux and spectral hardness (see SOM) is similar to that observed in blazars \cite{Takahashi96,Fossati00}. 

Accepting the accreting MBH hypothesis,  we now examine constraints on the BH mass and the accretion characteristics. We infer a minimum host-frame variability timescale of $t_{\rm var,min} \approx 78$ sec from the X-ray light curve (SOM). By requiring that $t_{\rm var,min}$ exceed the light-crossing time of $r_s$, we derive an upper limit on the MBH mass $M_{\rm BH} \simless 8\times 10^6 M_{\odot}$. Note that, like in gamma-ray burst light curves, even in the presence of relativistic motion (see below),  the observed variability should track that of the energy injection timescales from the central engine.  Irrespective of the timing argument, we can place approximate upper limits on the mass of the central BH if we assume the entirety of the galaxy mass [few$\times10^9 M_\odot$; \cite{levan2011}] and light [few$\times10^9 L_\odot$; \cite{levan2011}] arise from the host bulge (i.e., not in the disk) and apply the bulge mass--BH mass and the bulge luminosity--BH mass correlations \cite{mtr98,lfr+07,Kormendy11}.  All such analyses suggest $M_{\rm BH} \lesssim 10^{7}$\,$M_{\odot}$, securely under the limit (few$\times 10^8 M_\odot$) required for the tidal disruption of a solar-mass star to occur outside the event horizon of the MBH.

If the emission is isotropic, the average X-ray luminosity of the outburst (SOM), $L_{X} \approx 10^{47}$ erg s$^{-1}$, corresponds to the Eddington luminosity of a $\sim 10^9 M_{\odot}$ BH, incompatible with the upper limit derived from variability. If the source is relativistically beamed (see SOM), with beaming factor $f_b = (1 - \cos \theta_{\rm j}) \le 1$, the beaming-corrected luminosity $f_{b}L_{X} \sim 10^{45}$ erg s$^{-1}$ becomes consistent with the Eddington luminosity of a $\sim 10^{7}M_{\sun}$ SMBH if $\theta_{\rm j} = 1/\Gamma_{\rm j} \approx 0.1$, as inferred in blazars (we show below that this value of $\Gamma_{\rm j}$ is also consistent with the inferred rate of \ubs-like events).  We can also infer the presence of relativistic outflow (see SOM) by requiring that the true brightness temperature of the radio transient be less than the inverse Compton catastrophe temperature $10^{12}$\,K. Those constraints  require a mean $\Gamma_{\rm j} \age 1.9$ from $t_0$ to the time of our VLBI observations \cite{levan2011}.  Separately, we can use the observed variability of the radio counterpart to place constraints on the source size, finding $\Gamma_{\rm j} \gtrsim 10$.

\section*{A tidal disruption origin for the accretion mass}

If the source had been active in the distant past, we would expect to observe extended radio emission (e.g., jets or other emission knots) in our VLBI imaging. Since this is not seen and archival searches spanning two decades have yielded no evidence for prior AGN activity from radio to gamma-ray wavebands (see SOM), the evidence thus suggests that a $M_{\rm BH} = 10^6 - 10^7 M_\odot$ BH underwent a dramatic turn on to near-Eddington accretion rates, launching an energetic, relativistic outflow in the process. This rapid increase in the accretion rate cannot result from gas entering the sphere of influence (soi) of the MBH, since this would require a timescale $\sim R_{\rm soi}/\sigma \age 10^{4}$ yr to appreciably alter the accretion rate near the horizon, where $R_{\rm soi} \sim 1$ pc is the radius of the sphere of influence and $\sigma \sim 100$ km s$^{-1}$ is a typical bulge velocity dispersion. We suggest instead that a TDF provides a natural explanation for \ubs.

We again turn to the energetics and timescales to provide guidance on the TDF hypothesis. The observed X-ray fluence $S_X$ suggests an energy release of $E_X = 9.2 \times 10^{52} f_b$ erg for the first \hbox{$10^6$ s}. Assuming the source continues to radiate at a sustained level for $\sim$1 month, that the energy released in the XRT band is about 1/3 of the bolometric energy (following from Fig.\ 1) and adopting $f_b = 5\times 10^{-3}$, the total energy release from the jet amounts to 0.3\% of the mass-energy accreted if $M_* = M_\odot$. Given a typical accretion efficiency of $\epsilon_{\rm BH} \equiv E_{\rm av}/m_{\rm acc}c^2= 0.1$, the jet need radiate only about 1/30th of the available energy $E_{\rm av}$. The duration of the X-ray light curve and the requisite accretion rate are also compatible with the several-day fallback timescale (see SOM).

The broadband SED of \ubs\ shown in Fig.\ 1 displays two peaks, at far infrared and X-ray/gamma-ray wavebands, respectively. The overall spectral shape is reminiscent of blazars, for which the peaks at low and high energies are typically modeled as synchrotron and Inverse Compton (IC) emission, respectively.  The X-ray emission shows both a bright/flaring and a dim/slower-varying (``quiescent'') state.  Under the TDF hypothesis, what could account for the observed spectrum and temporal behavior?

\begin{itemize}
\item {\bf Single Component Synchrotron with Dust Extinction:} In low-luminosity BL Lac objects, the $\nu F_{\nu}$ synchrotron spectral peak may occur at energies as high as hard X-rays.  Thus, one possibility is that the entire emission from radio to X-rays is part of a single non-thermal synchrotron spectrum originating from shocked relativistic electrons.  In this scenario, the suppressed optical emission and red IR colors of the transient could result from dust extinction with $A_V > 10$ mag. We conclude that although a single extinguished synchrotron spectrum cannot be ruled out, the large required extinction may disfavor this interpretation (see Fig.\ S3).  Furthermore, although a synchrotron origin is still likely for at least the radio emission, there is evidence that the radio and X-ray emitting regions may not be coincident (see below).

\item {\bf Two-Component Blazar Emission:} The FIR and hard X-ray peaks may, instead, represent distinct spectral components, corresponding to synchrotron and IC emission, respectively, as in blazars.  Figure 1 shows that the $\nu F_{\nu}$ luminosity of the low energy peak is $\sim 1-2$ orders of magnitude weaker than the high energy peak.  This extreme ratio, and the relatively low frequency of the synchrotron peak, are both compatible with Eddington-accreting blazar emission \cite{Fossati+98}.  We consider various origins for the seed photons for IC in the SOM and show two example model SEDs in Figure 1.

\item {\bf Forward Shock Emission from Jet-ISM Interaction:} Although the above models generally assume that the low- and high-energy spectral components are directly related, evidence suggests that they may originate from distinct radii, at least during the X-ray flaring state.  While the rapid variability of the X-ray emission strongly indicates an ``internal'' origin \cite{2001MNRAS.325.1559S}, the radio-IR emission varies more smoothly and could instead result at larger radii from the interaction of the jet with the surrounding interstellar medium (see SOM).  If no AGN activity occurred prior to the recent onset of emission, the jet must burrow its way through the gas in the nuclear region.\footnote{This situation is not encountered in normal (long-lived) blazars because a large $\gtrsim$ kpc scale cavity has been carved by the preceding outflow.}  Due to its fast motion, the newly-formed jet drives a shock into the external gas (forward shock), while simultaneously a reverse shock slows it down.  Particles accelerated at these shocks may power synchrotron afterglow emission beginning simultaneously when the jet forms, yet lasting long after the internal emission has faded.  This model, the geometry of which is depicted in Figure 2, appears to best accommodate the data, and makes specific predictions for the long-term fading of the radio and IR transient (see SOM).  
\end{itemize}

\section*{Conclusions and Predictions}

No rising UV-optical transient nor slowly evolving thermal X-ray component has been seen to date from \ubs, in contrast with the nominal expectations of TDFs. However, if \ubs\ was obscured by dust, then UV-optical suppression of the transient would be natural. And if we understand the thermal X-ray emission as being outshone by the jetted emission in the first week, then the TDF hypothesis would naturally lead to the prediction of the emergence of the thermal component on timescales of weeks to months.  Whether and when thermal emission will be observable hinges on the degree of dust extinction, its brightness relative to the host bulge, and how rapidly the jet emission fades.

If the TDF hypothesis is correct, \ubs\ will fade over the coming year and will not repeat. If our interpretation about the relativistic flow and spectral origin is correct, then we would expect the transient emission to be polarized at a (low) level similar to that seen in gamma-ray burst afterglows (as opposed to blazars\footnote{Here the departure from the blazar analogy is worth noting in that the physics of the radio emission is likely to be different in this case: we have suggested that the emission is originating from the shocked surrounding material (forward shock), not the shocked jet as in normal blazars, which could contain large scale fields. Even so, only 10\% of flat-spectrum  radio quasars and BL Lac objects have polarization larger than our VLBI limits \cite{2003ApJ...586...33A}.}). Moreover we expect to see evidence for superluminal motion of the radio source as seen in VLBI monitoring over the next few months; the source itself may become resolved on timescales of a few months if it remains bright enough to detect at radio wavebands.

Adopting a beaming fraction $f_{b} \lesssim 10^{-2}$ consistent with that inferred from \ubs\ (SOM), we conclude that for every on-axis event, there will be $1/f_{b} \gtrsim 10^{2}$ events pointed away from our line of sight.  Since Swift has triggered on one event in $\sim 6$ years of monitoring, the total inferred limit on the rate of TDFs accompanied by relativistic ejecta is $\gtrsim 10$ yr$^{-1}$ out to a similar distance.  Although the majority of such events will not produce prompt high-energy emission, bright radio emission is predicted once the ejecta decelerates to non-relativistic speeds on a timescale $\sim 1$ year \cite{2011arXiv1102.1429G}.  The predicted peak flux is sufficiently high ($\sim 0.1-1$ mJy at several GHz frequencies and redshifts similar to \ubs) that $\sim 10-100$ may be detected per year by upcoming radio transient surveys.  Continued long-term monitoring of the radio afterglow from \ubs\ will allow for calibration of the off-axis models. 

We conclude with two broader ramifications that stem from our understanding of the origin of \ubs. The emerging jet from the tidal disruption event appears to be powerful enough to accelerate cosmic rays up to $\sim 10^{20}$ eV, i.e., the highest observed energies \cite{FarrarGruzinov09}. The observed rate of jets associated with the tidal disruption of a star, $\dot{R}\sim 10^{-11}$Mpc$^{-3}$yr$^{-1}$, and the energy released per event of $E_X\sim 10^{53}$ erg, however, imply an energy injection rate of $\dot{E}_{\rm TDF}\sim 10^{42}$ erg Mpc$^{-3}$yr$^{-1}$. Despite the large uncertainty, this rate is significantly smaller than the injection rate $\dot{E}_{\rm inj}\sim 10^{44}$ erg Mpc$^{-3}$yr$^{-1}$ required to explain the observed flux of cosmic rays of energy $> 10^{19}$ eV \cite{Waxman95}.  This conclusion is, however, subject to uncertainties associated with the radiative efficiency of the jet. 

There is much evidence that AGN jets are accelerated by magnetohydrodynamic, rather than hydrodynamic, forces \cite{2004ApJ...605..656V}.  A key unsolved question is whether the large-scale magnetic field necessary to power the jet is advected in with the flow \cite{2005ApJ...629..960S}, or whether it is generated locally in the disk by instabilities or dynamo action.  If the jet is launched from a radius $R_{\rm in}$, the magnetic field strength at its base ($B$) is related to the jet luminosity by $L_{\rm j} \sim \pi R_{\rm in}^2 c\times (B^2/4\pi)$.  If we assume $L_{\rm j} \sim 10^{45}$ ergs s$^{-1}$, similar to the Eddington limit for a $\sim 10^7 M_{\odot}$ MBH (as appears necessary to explain the bright non-thermal emission), the required field strength is $B \sim 10^{5}$ G for $R_{\rm in} \sim 1.5 r_s$.  This field is much higher than the average field strengths of typical
main sequence stars ($< 10^{3}$ G). The stellar field is further diluted due to flux freezing by a factor $\sim (R_*/R_{\rm in})^{2}$ as matter falls into the BH, where $R_{\star} \sim R_{\odot}$ is the stellar radius prior to disruption.  Hence we conclude that the large-scale field responsible for launching the jet associated with \ubs\ must have been generated in situ.  Placing similar constraints has not previously been possible in the context of normal AGN or X-ray binary disks, due to the much larger ratio between the outer and inner disk radii in these systems.  

%

\begin{scilastnote}

\item We thank R. Romani, C. McKee and L. Blitz for close reads of drafts of this work and for helpful interactions. We are grateful to entire Swift team for work on their remarkable facilities that enabled discovery of this event. Swift, launched in November 2004, is a NASA mission in partnership with the Italian Space Agency and the UK Space Agency. Swift is managed by NASA Goddard. Penn State University controls science and flight operations from the Mission Operations Center in University Park, Pennsylvania. Los Alamos National Laboratory provides gamma-ray imaging analysis. JSB and his group were partially supported by grants NASA/NNX10AF93G, NASA/NNX10AI28G, and NSG/AST-100991. DG acknowledges support from the Lyman Spitzer, Jr.\ Fellowship awarded by the Department of Astrophysical Sciences at Princeton University. 
BDM is supported by NASA through Einstein Postdoctoral Fellowship grant number PF9-00065 awarded by the Chandra X-ray Center, which is operated by the Smithsonian Astrophysical Observatory for NASA under contract NAS8-03060. SBC wishes to acknowledge generous support from Gary \& Cynthia Bengier, the Richard \& Rhoda Goldman fund, NASA/Swift grant NNX10AI21G, NASA Fermi grant NNX10A057G, and NSF grant AST-0908886. WHL is supported in part by CONACyT grant 83254.  
AJvdH was supported by NASA grant NNH07ZDA001-GLAST.
\end{scilastnote}

\noindent \includegraphics[width=6.25in]{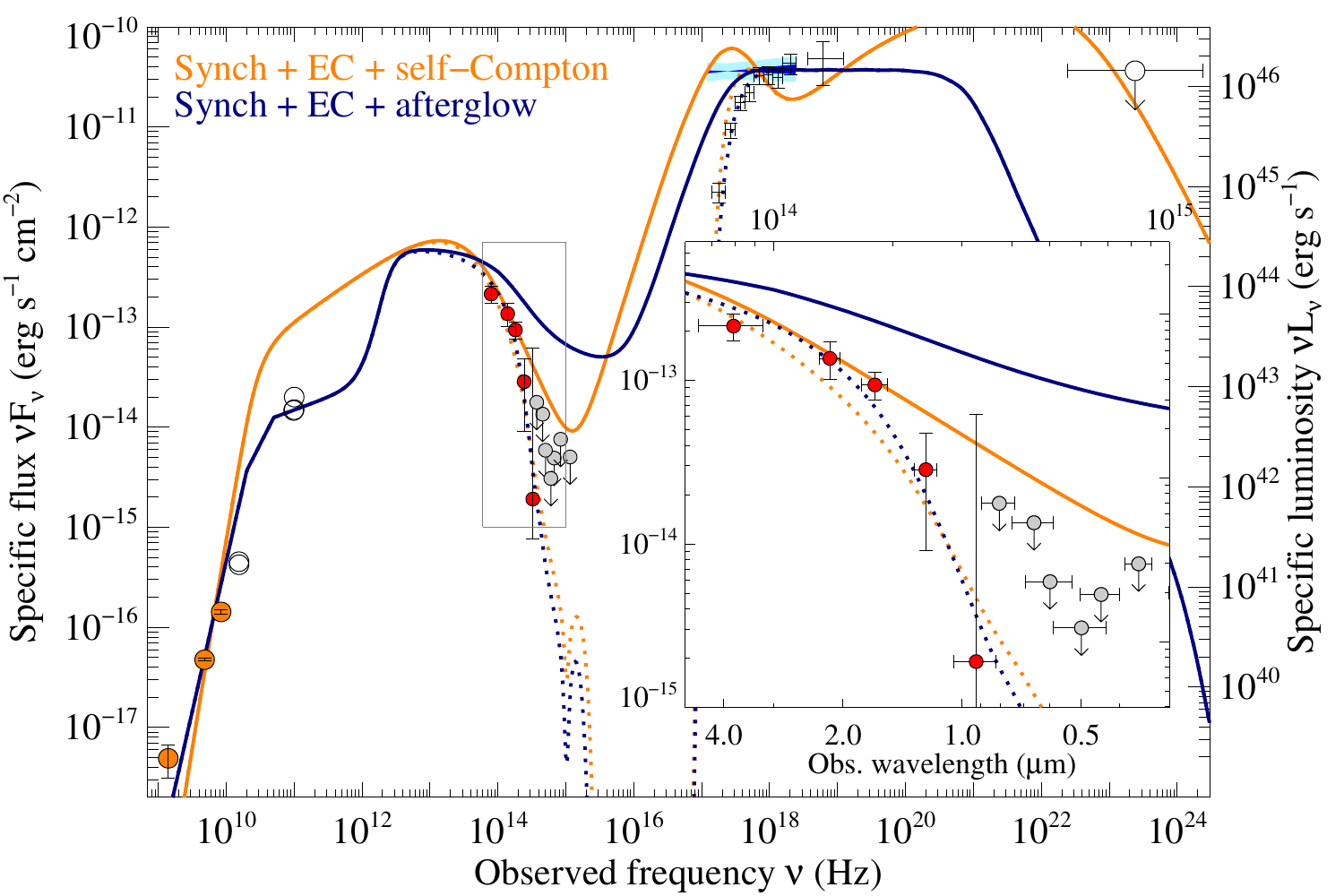}\\
\noindent {\bf Fig.\ 1: Multiwavelength spectral energy distribution of \ubs\ at $t_0$ + 2.9 day.  Our radio-through-UV measurements are represented by filled circles, with data from the published circulars \cite{levan2011} represented by open circles; the uncertain relative contributions of the host galaxy and the optical/infrared (IR) transient result in very large uncertainties for the $J$ and $z$ photometric data points. X-ray and soft gamma-ray points from the Swift XRT and BAT  (uncorrected for host-galaxy absorption) are shown as black crosses, and the Fermi/LAT gamma-ray upper-limit \cite{fermi2011} is shown at the far right.  The 90\% uncertainty region of a power-law fit to the XRT data (with $N_H$ absorption removed) is represented by the blue bow-tie.  The inset at lower right shows the same data zoomed in on the optical-NIR window.
These observations are overplotted with two different multi-component models for the SED, exploring different emission mechanisms and radii (see also Fig.\ 2). The orange curve shows a model with synchrotron, synchrotron self-Compton, and external Compton contributions.  In this model, the radio and IR emission are produced by synchrotron radiation from an extended source, while the X-ray emission is dominated by the Compton scattering of external photons from the accretion disk (for illustrative purposes, we assume a $10^{6}M_{\odot}$ MBH).  The purple curve shows a model in which the IR emission originates from a compact source of synchrotron emission, as may be required by the IR variability ($\sim4 \times 10^{14}$ cm).  As in the orange model, the X-ray emission is dominated by external Compton scattering, while the peak at high energies results from synchrotron self-Compton emission.  An additional synchrotron component from a mildly relativistic blast-wave afterglow at larger radius is invoked to explain the bright radio and millimeter fluxes.  Both models require moderate extinction ($A_V \sim 3-5$ mag).  Additional synchrotron models are shown in Figure S3.  The model SEDs in this Figure and in Figure S3 were generated using the computer code from Ref.~\cite{Krawczynski+04}. }

\newpage

\noindent \centerline{ \includegraphics[width=3.8in]{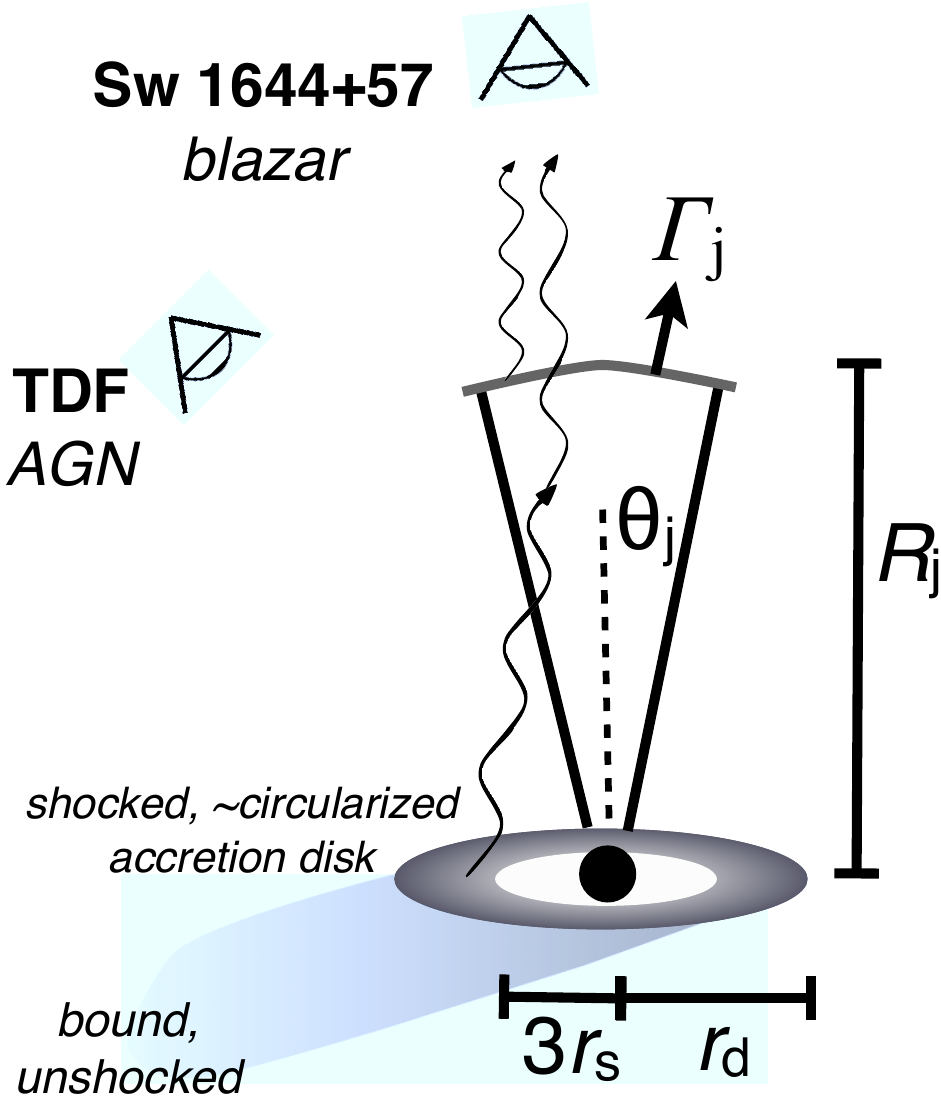}\\}
\noindent {\bf Fig 2. Schematic representation of the geometry and emission regions for \ubs. A star is disrupted at distance $r_{\rm d}$ from a black hole of mass $M_{\rm BH}$ with Schwarzschild radius $r_{\rm s}$. Half of the mass of the star escapes on unbound orbits while the other half remains bound. Shocked, circularized fallback mass sets up a temporary accretion disk with inner radius $3 r_{\rm s}$ (for a non-spinning BH). A two-sided jet is powered starting at the time of accretion and plows through the interstellar region surrounding the BH at a Lorentz factor $\Gamma_{\rm j}$. At some later time, the jet has reached a distance $R_{\rm j}$ where the forward shock radiates the observed radio and infrared light. Emission from the accretion disk is Compton upscattered giving rise to the observed X-rays. Different viewing angles (whether the observer is inside $\theta_{\rm j} \approx 1/\Gamma_{\rm j}$ or not) determines what sort of phenomena is observed. An analogy with blazars and AGN for more massive BHs is given.}

\newpage 


\section*{Supplementary material}

# 1 Supplemental Online Material

"A relativistic jetted outburst from a massive black hole fed by a tidally disrupted star," by J. S. Bloom et al.

In what follows, we assume cosmological parameters $H_0 = 73 \, \text{km s}^{-1} \text{Mpc}^{-1}$, $\Omega_\Lambda = 0.73$ and $\Omega_m = 0.27$. At a redshift of $z = 0.3534$ (1), the luminosity distance is 1814.8 Mpc and the angular diameter distance $d_A = 990.8$ Mpc (with 1 arcsec corresponding to 4.803 kpc in projection).

## 1.1 Timing and Energetics Analysis of the X-ray Light Curve

Figure S1 shows the X-ray light curve in the first 16.6 days from the Swift/BAT trigger at time $t_0$. The total on-source integration time of the XRT was $\sim 3$ d. From the XRT light curve we can infer a minimum timescale of variability over which the flux changes by a factor of 2. We find this minimum observed $t_{\text{var,min}} = 105$ sec occurs as a rise between $t_0 + 111499$ and $t_0 + 111604$ sec. The most rapid observed decline ($t_{\text{var}} = 188$ s) occurred much later, at a time $t_0 + 1418381$ and $t_0 + 1418569$ sec, as the XRT light curve began at major secular downturn. From a periodogram analysis, we also (2, 3) find no significant frequencies in the XRT light curve that are not associated with the orbital period of the spacecraft.

As in accreting BH systems the X-ray hardness tracks the X-ray flux of Sw 1644+57 (4, 5), such that the best-fit powerlaw index increases with decreasing flux (1). From structure function analysis (Fig. S2), we measure a red-noise power spectrum, characterized by a powerlaw slope with frequency $\alpha = -1.6 \pm 0.2$. This is consistent with the typical value [$\alpha = -2.0 \pm 0.5$; (6)] at X-ray wavebands for blazars. Such a red noise spectrum is also typical of low-frequency ($f < 1$ Hz) X-ray variability in BH and neutron star binaries (7). We thus interpret the power density spectrum of the X-rays as indicative of accretion around a central compact mass.

The peak luminosity observed in the X-rays is $L_{\text{X,peak}} \sim 3.0 \times 10^{48}$ erg s$^{-1}$ and the average over the first $1.2 \times 10^6$ seconds is $L_X \approx 9.2 \times 10^{46}$ erg s$^{-1}$. The fluence within 0.3 keV to 10 keV—which is $\gtrsim 20$ times that from longer wavelengths (see Fig. 1 of the main text and Fig. S3)—from $t_0$ to $t_0 + 1.2 \times 10^6$ s is $S_X = 3 \times 10^{-4}$ erg cm$^{-2}$.

## 1.2 Archival Search for AGN activity

One possible explanation for the observed activity is that Sw 1644+57 represents an unusual AGN that went undetected up to this point. However, archival Swift observations place constraints on existing bright gamma-ray and X-ray emission dating back to Feb. 2005 (*8*), with typical one-day average luminosities $L_{\rm X,pre} \lesssim 1.7 \times 10^{44}$ erg s$^{-1}$, a factor of about $10^3$ less than the average luminosity of Sw 1644+57. The source was also not detected in the ROSAT All-Sky Survey. The upper limit of $1.7 \times 10^{44}$ erg s$^{-1}$ implies no similar activity back to 1990 (Pilar Esquej, private communication). Furthermore, the FIRST survey places a 1.4 GHz (3 $\sigma$) upper limit of $\lesssim 0.3$ mJy on quiescent radio emission (*9*). We know of no pre-event spectrum of the host galaxy, but post-event spectroscopy analysis showed line-ratio characteristics of a star-forming galaxy without any indication of AGN activity (*1*).

## 1.3 Inference of Relativistic Outflow & Beaming

**High-Energy Constraints**: Beaming of the observed X-ray emission (*10*) could be due either to purely geometric effects associated with the accretion disk and its outflows (*11, 12*) or due to relativistic effects in a collimated jet. Although both possibilities are plausible *a priori*, the presence of bright non-thermal radio emission strongly suggests the presence of some relativistic material. Geometrically-beamed emission could be observed, even if the observer is not directly along the line of sight of the relativistic jet. However, the fact that the beaming fraction $f_b \simeq \theta_{\rm j}^2/2 \sim 5 \times 10^{-3}$ of typical blazar jets (with $\theta_{\rm j} \sim 0.1$, the half opening angle; Fig. 2 main text) is similar to the beaming estimated from the predicted TDE rate (§1.5) suggest that this possibility is not essential. The beaming-corrected luminosity in this case $f_b L_X \sim 10^{45}$ erg s$^{-1}$ is also then consistent with the Eddington luminosity of a $\sim 10^7 M_\odot$ MBH.

Another possible explanation for very bright X-ray emission is that the accretion rate, and the resultant jet power, is above the Eddington rate. Although super-Eddington accretion is indeed predicted at early times in TDE associated with low SMBH masses and super-Eddington jets may have empirical support (*13*), completely unbeamed emission is inconsistent with the rate estimates (§1.5).
**Radio Constraints**: We can constrain the outflow velocity of Sw 1644+57 by making use of the well-known brightness temperature limits on synchrotron radi-

2

ation (*14, 15*). The (frequency-dependent) brightness temperature is given by:

$$T_{\rm B}(\nu) = \frac{c^2}{2k_{\rm B}\nu^2}\left(\frac{S_\nu}{\pi\theta^2}\right). \qquad (1)$$

Here $S_\nu$ is the observed flux density at frequency $\nu$, $\theta$ is the angular size of the source, and $k_B$ is Boltzmann's constant. At brightness temperatures above $\gtrsim 10^{12}$ K, multiple inverse Compton scattering of synchrotron radio photons to $\gamma$-ray and X-ray energies begin to rapidly dominate the luminosity. This is known as the "inverse Compton catastrophe", and effectively determines a maximal brightness temperature for a radio synchrotron source.

Our most direct constraint comes from our VLBI detection on 2011 April 1 (*1*), where we have measured $S_\nu(\nu = 8.5\,{\rm GHz}) = 1.7\,{\rm mJy}$. If we assume $T_{\rm B} \lesssim 10^{12}$ K, we infer the angular size of the radio emitting region must be $\theta \gtrsim 3\,\mu{\rm as}$, or $r \equiv \theta d_{\rm A}/2 \gtrsim 3 \times 10^{16}$ cm. Assuming that the outflow began at the time of the initial *Swift* detection of Sw 1644+57 (2011 March 25), this corresponds to a mean apparent velocity of $\langle v \rangle \equiv r/\Delta t \gtrsim 1.2c$ (ie., mild superluminal expansion). We note that this is entirely compatible with the fact that we have yet to (spatially) resolve the radio emission from Sw 1644+57: the upper limit on the angular size of $\theta < 1\,{\rm mas}$ corresponds to an upper limit on the mean apparent velocity of $\langle v \rangle \lesssim 400c$.

For relativistic sources, Equation 1 must be modified to account for the difference between properties measured in the rest frame of the emitting fluid and by an observer at rest with respect to the outflow. Accounting for the appropriate Doppler factors as described in ref. (*15*) (their Equation 3), our VLBI detection requires $\Gamma \equiv 1/\sqrt{1-\beta^2} \gtrsim 1.9$; in other words, the outflow must be at least mildly relativistic.

Separate from the above arguments based upon our VLBI measurement, we can use the observed variability of the radio counterpart to place constraints on the source size. The radio transient varies by approximately a factor of two over a time span of $\delta t \approx 7$ hours in observations reported by the Owens Valley Radio Observatory [OVRO; (*16*)], and by $\approx 50\%$ over a time span of $\delta t \approx 6$ hours from data obtained at the Mullard Radio Astronomy Observatory [MRAO; (*17*)]. We conservatively adopt a variability timescale of 1 day for the radio (15 GHz) flux to change significantly. This variability could either be due to interstellar scintillation (ISS), as has been observed in the radio afterglows of gamma-ray bursts (*18*), or could be intrinsic to the source.

In order to observe such strong variability due to scintillation, the observed radio bands would need to be in the strong, diffractive regime, where the coherence

3

length scale is smaller than the Fresnel scale. This requires:

$$\nu < \nu_{ss} \equiv 13.4 \left(\frac{\text{SM}}{10^{-3.19}\text{m}^{-20/3}\text{kpc}}\right)^{6/17} \left(\frac{D_{scr}}{\text{kpc}}\right)^{5/17} \text{GHz} \qquad (2)$$

Here SM is the electron scattering measure, and $D$ is the distance to the scattering screen. Using the electron distribution model of (*19*), we infer SM = $10^{-3.66}$ m$^{-20/3}$ kpc. For typical scattering screen distances ($D \approx 0.1$–1 kpc), the scintillation frequency is thus $\nu_{ss} \lesssim 9$ GHz. It seems likely, then, that the 15 GHz observations are not affected by strong ISS. However, we caution that the scattering measure is only coarsely mapped, and small scale features can lead to large variations in SM.[1]

If the variability is indeed intrinsic to the source, we can use the observed timescale to independently constrain the source size. Based on causality arguments, the emitting region must be smaller than the light crossing time, and so $\theta \lesssim \Gamma c \delta t / d_A \approx 0.2\Gamma$ $\mu$as) at $\nu = 15$ GHz. Associating this angular size with a (relativistic) brightness temperature, we find $\Gamma \gtrsim 10$, broadly consistent the limits on the angular sized and Lorentz factor inferred from the VLBI measurement[2].

## 1.4 Timescales and Accretion Rates for TDFs

After the star is disrupted, a sizable fraction of its mass is placed onto highly eccentric orbits, which return it to the vicinity of the black hole on a timescale (*22*):

$$\begin{aligned} t_{\text{fallback}} &\simeq \frac{2\pi}{6^{3/2}} \left(\frac{R_p}{R_\star}\right)^{3/2} \left(\frac{R_p^3}{GM_{\text{BH}}}\right)^{1/2} \\ &\simeq 5 \left(\frac{M_{\text{BH}}}{10^7 M_\odot}\right)^{5/2} \left(\frac{R_p}{6GM_{\text{BH}}/c^2}\right)^3 \left(\frac{R_\odot}{R_\star}\right)^{3/2} \text{ days}, \end{aligned} \qquad (3)$$

where in the last expression we consider a solar-type star ($R_\star = R_\odot$) with a perinigricon distance[3] $R_p = 6GM_{\text{BH}}/c^2$. The fall-back accretion rate peaks at

---

[1] We note that there are no known supernova remnants (presumably arising in a particularly dense region of the ISM) in the vicinity of Sw 1644+57 (*20*).

[2] We caution that such causality arguments have difficulty explaining the observed fluctuations from the so-called "inter-day variables" (e.g., (*21*)) without resorting to extreme Lorentz factors ($\Gamma \gtrsim 100$), and so may not necessarily be applicable to Sw 1644+57.

[3] We are considering here main sequence stars since giants will disrupt at larger radii and hence produce smaller accretion rates and (likely) less luminous jets.

$t \sim t_{\text{fallback}}$ and declines $\propto t^{-5/3}$ thereafter (*23, 24*). Note that $t_{\text{fallback}} \sim$ days sets a characteristic timescale for the onset of bright emission, consistent with the observed onset of the activity associated with Sw 1644+57.

The above estimates ignore the detailed physics of energy release near the MBH, assuming that the observed luminosity indeed tracks the instantaneous mass fallback rate. Even completely circularized accretion flows may be susceptible to viscous and thermal radiation pressure-driven instabilities at high accretion rates (e.g., (*25*)). However, this is unlikely to be responsible for the short timescale variability, because the viscous timescale, $t_{\text{visc}} \sim \alpha^{-1}(r/h)^2 \Omega^{-1}$ is significantly longer than the dynamical time, $\Omega^{-1}$, unless both the disc viscosity parameter, $\alpha$, is close to unity, and the disc geometrical aspect ratio is high, $h/r \sim 1$. A more likely source of short, dynamical timescale variability is the initial circularization of the disrupted material, which initially has high orbital eccentricity (*26–28*). The presence of the jet suggests that the BH may be rapidly spinning; thus there may be significant orbital precession near the horizon due to general relativistic effects. This precession could cause the falling-back stream of bound gas not to intersect itself on its first passage through perinigricon, instead taking several orbital periods to shock and fully circularize its orbit (*26*); perhaps partial shocking and accretion at each pericenter passage gives rise to the observed X-ray variability.

One plausible (though hardly unique) scenario that broadly describes the observed lightcurve is as follows: [1] $t \lesssim 10^5$ s: initialization of a highly irregular jet; [2] $t \sim 3 \times 10^5$ s: disc circularization phase, during which several large flares are powered as the bulk of the disrupted material on an eccentric orbit passes through the perinigricon (which probably strongly modulates the jet power and the supply of seed photons); [3] longer-term viscous evolution of the disk, characterized by a steadier consumption of material, fed by the continuing fallback of the disrupted stellar material, as described by equation 3; [4] a decrease in the accretion rate from fallback material, leading to the dimming of the transient. Note that in this scenario the characteristic variability timescale is predicted to increase during the circularization phase. Circularization is probably mediated by strong shocks and may require up to several several local orbital times to complete. Thus, material closest to the SMBH circularizes first, while material further out requires a correspondingly longer time. This behavior is qualitatively consistent with the observed X-ray lightcurve.

## 1.5 Rate Estimates under the TDF Hypothesis

We can obtain a model independent estimate of beaming by comparing the detection rate of TDF events by Swift with the estimated occurrence rate. Assuming a tidal disruption rate per galaxy $R_{\rm TDF} \equiv R_{-5} 10^{-5}$ yr$^{-1}$ and adopting a local MBH density $\sim 10^{-3} - 10^{-2}$ Mpc$^{-3}$ (e.g. (*29*)), the predicted TDF rate out to the redshift $z = 0.3534$ (co-moving volume $\simeq 11$ Gpc$^3$) is $\sim 10^2 - 10^3 R_{-5}$ yr$^{-1}$. Since Swift has triggered on only one TDF in $\approx 6$ years of continuous monitoring[4], the required beaming fraction is $f_b \sim 2 \times 10^{-3} - 2 \times 10^{-4} R_{-5}^{-1}$. Adopting typical values for $R_{-5} \sim 1 - 10$ predicted by theoretical studies (*31, 32*) and consistent with the TDF rate estimates from ROSAT (*33*), XMM-Newton (*34*), and GALEX (*35*) and constraints on off-axis radio events (*36, 37*), we find that $f_b \sim 10^{-4} - 10^{-2}$. As discussed in the main text, this beaming is broadly consistent with the typical opening angles of blazar jets. This simple estimate above may need to be altered if the disruption rate or SMBH density is higher than our adopted fiducial range [as may be the case if the galaxy is small and compact and/or underwent a recent merger; (*38, 39*)]; or if only a fraction of on-axis TDF events produce nonthermal emission [as may be the case if another parameter other than the accretion—such as the SMBH spin—is necessary for jet production; e.g. (*40*)].

## 1.6 Argument for Separate Radio and X-Ray Emission Radii

We now show that the radio emission and the X-rays (at least the variable contribution) are unlikely to originate from the same spatial location. This lends credence to the suggestion in the main text that the X-rays are of "internal" origin, whereas the radio (and possibly IR) emission may originate from larger radii in the forward shock afterglow.

If the emitting material has a Lorentz factor $\Gamma_j$, then from the minimum host-frame variability timescale $t_{\rm var,min} \sim 78$ s, we infer the characteristic radius of the X-ray emitting region:

$$R_X \sim c\Gamma_j^2 t_{\rm var,min} \approx 2.3 \times 10^{14} \Gamma_{10}^2 \text{ cm}, \qquad (4)$$

where $\Gamma_j \equiv 10\Gamma_{10}$.

Separately, the radius of the radio-emitting region can be derived using the fact that any (incoherent) emission process cannot be brighter than the blackbody

---

[4]Note, however, that given the detection significance of the image trigger and the brighter later peaks in the light curve (*30*), this transient would have been eventually detected by Swift/BAT within a larger volume.

6

(BB) flux. Hence we demand that

$$F_{\nu_{\rm obs}} < F_{\nu_{\rm obs},BB} = \frac{L_{\nu_{\rm obs},BB}}{4\pi D_{\rm L}^2} = \frac{32\pi}{3}\frac{R^2}{D_{\rm L}^2}\nu_{\rm obs}^2\gamma_{\rm e}m_{\rm e}(1+z)^3 \quad (5)$$

where

$$L_{\nu_{\rm obs},BB} = \frac{16\pi}{3}r^2c\Gamma_{\rm j}^2 u_\nu, \quad (6)$$

is the specific BB luminosity and

$$u_\nu = \frac{8\pi\nu^2 kT_{\rm e}}{c^3} = \frac{8\pi\nu_{\rm obs}^2 kT_{\rm e}}{c^3\Gamma_{\rm j}^2} \quad (7)$$

is the rest-frame BB energy density, $\nu(\nu_{\rm obs})$ are the rest-frame (observer) frequencies, and $kT_{\rm e} \approx \gamma_{\rm e} m_{\rm e} c^2$ is the "temperature" of the radiating electrons, corresponding to the characteristic (minimum) random Lorentz factor of the emitting electrons $\gamma_{\rm e}$.

If the radio emission is synchrotron, the peak frequency occurs at

$$\nu_m = \frac{eB\gamma_{\rm e}^2\Gamma_{\rm j}}{2\pi m_{\rm e} c}, \quad (8)$$

where $B$ is the magnetic field strength in the emission region. If we assume that a fraction $\epsilon_{\rm B}$ of the isotropic jet luminosity $L_{\rm j}$ is carried by Poynting flux, then the field strength at radius $r$ is given by

$$B^2 = \frac{\epsilon_{\rm B} L_{\rm j}}{r^2 c\Gamma_{\rm j}^2} \quad (9)$$

Combining equations (8) and (9) we solve for the electron Lorentz factor

$$\gamma_{\rm e} = 10^2 L_{\rm j,47.5}^{-1/4} r_{15}^{1/2} \nu_{m,13}^{1/2} \epsilon_{\rm B,-2}^{-1/4}, \quad (10)$$

where $L_{\rm j} \equiv 10^{47.5} L_{\rm j,47.5}$ ergs s$^{-1}$, $r \equiv 10^{15} r_{15}$ cm, $\epsilon_{\rm B} = 10^{-2}\epsilon_{\rm B,-2}$, and $\nu_m \equiv 10^{13}\nu_{m,13}$ Hz.

Now using equation (10) with (5) to constrain the emitting radius

$$r \gtrsim 1.2 \times 10^{16} F_{\nu_{\rm obs},\rm mJy}^{2/5} \nu_{10}^{-4/5} L_{\rm j,47.5}^{1/10} \nu_{m,13}^{-1/5} \epsilon_{\rm B,-2}^{1/10}\,{\rm cm}, \quad (11)$$

where $F_{\nu_{\rm obs}} = F_{\nu_{\rm obs},\rm mJy}$mJy, and $\nu_{\rm obs} \equiv 10\nu_{10}$ GHz.

From the $\sim 1.7$ mJy VLBI detection at $\nu_{10} = 0.85$, equation (11) shows that the emitting radius $r \gtrsim 2 \times 10^{16}$ cm, rather insensitive to the precise values of the

7

jet luminosity, synchrotron peak frequency, and $\epsilon_{\rm B}$. This constraint agrees with the minimum radius derived in §1.3, but remains valid even for large values of $\Gamma_{\rm j}$. *Note that the minimum radius of the radio-emitting plasma is incompatible with the X-ray radius $R_X$ (eq. [4]) unless $\Gamma_{\rm j} \gtrsim 100$, significantly higher than values typically inferred for AGN jets.* If the jet Lorentz factor were indeed this high, it would furthermore imply that the cooling timescale of the X-ray emitting electrons is much longer than $t_{\rm var,min}$, a contradiction (*41*). We conclude that the radio emission and the X-rays (at least during the flaring state) originate from distinct radii.

## 1.7 Origin of Inverse Compton Seed Photons in the Two Component Model

One possible source of up-scattered photons in the two-component model is low-energy synchrotron emission itself (synchrotron self-Compton; SSC). In Figure 1 of the main text we show model SEDs for the quiescent phase which include synchrotron-SSC emission. In general we find that pure synchrotron-SSC models, in which the source originates from sufficiently compact radii to explain the IR/X-ray variability, cannot by themselves explain the large ratio between the X-ray and IR luminosities. In addition, the peak flux of the second Compton peak is only marginally consistent with the *Fermi* upper limits.

A more promising source for the X-ray emission (especially during the flaring phase) is *external IC*, as is thought to characterize the brightest blazars (e.g. (*42*)). Since this can in principle occur at smaller radii near the SMBH, it may also accommodate the flaring. What is the origin of the external photons? On timescales of just a few days, emission from the disk has not had sufficient time to establish a Broad Line Region (which is generally expected to form at $r \sim 10^{17}$ cm). Nevertheless other sources of soft photons exist within a distance $r \sim 50 r_{\rm s} \sim 10^{14}$ cm. One source is the thermal emission from the accretion disk, which is characterized by a temperature $\sim 10 - 100$ eV and luminosity $\sim L_{\rm edd} \sim 10^{44-45}$ erg s$^{-1}$. However, for the jet to efficiently scatter the disk flux, then full acceleration to $\Gamma_j \sim 10$ must occur within a distance $\sim 5 - 10 r_{\rm s}$, since otherwise the disk photons originate from the back and do not provide an efficient source (e.g., (*43*)).

Another source of external photons is the shock where the bound gas falls back to perinigricon. While the fallback rate is super-Eddington, radiation pressure at the shock likely drives a powerful outflow of gas (*11, 44*) that can extend out to $\sim 50 r_{\rm s}$, a distance sufficient to extract sufficient energy from the jet via external

8

IC to power the observed X-rays. If the density at the shock is high enough to produce many photons, the radiation reaches thermal equilibrium and peaks in the optical/UV. IC scattering by even cold electrons in the jet will boost UV photons by a factor of at least $\Gamma_j^2 \sim 100$ (or higher if the scattering electrons are hot), resulting in hard X-ray emission. However, for $M_{\rm BH} = 10^7 M_\odot$, the density in the shock may be too low for the radiation to reach thermal equilibrium; in this case, IC in the vicinity of the shock produce photons with energies $\gtrsim 10 - 100s$ keV (*44*). These could then be further upscattered by the jet. We conclude that external IC is also a viable explanation for the X-ray emission.

## 1.8 Forward Shock Emission Predictions

Adopting the X-ray emission as a proxy for the jet power and duration, we assume that the bulk of the energy is ejected over a timescale $T_{\rm ej} \sim 2$ days, with average isotropic-equivalent luminosity $L_j \equiv 3 \times 10^{47} L_{47.5}$ erg s$^{-1}$. The total energy release during this initial period is $E_{\rm iso} \sim 10^{53}$ erg, after which the X-ray luminosity drops by a factor of several.[5] At a large distance, we model the jet as a homogeneous shell of thickness $\delta = cT_{\rm ej} \equiv \delta_{16} 10^{16}$ cm, which propagates into an external medium of density $n_{\rm ext}(r)$. The circumnuclear gas is expected to be rather dense out to the Bondi radius $R_b$ due to e.g. gas accretion (e.g. (*45*)). For simplicity, we assume that the density scales $\propto r^{-2}$ for $r < R_b = 0.1$ pc, but then smoothly connects onto a constant density profile $n_{\rm ext} = 10 n_{10}$ cm$^{-3}$ further out. Our results depend rather weakly on the details of the profile as long as $n_{\rm ext} \sim 10$ cm$^{-3}$ for $r \gtrsim 0.1$ pc.

The interaction between the jet and the external medium results in a characteristic forward/reverse shock structure. The Lorentz factor of the shocked fluid is (*46*)

$$\Gamma_{\rm sh} = \left(\frac{\Gamma_j^2 n_j}{4 n_{\rm ext}}\right)^{1/4} = 3.4 L_{47.5}^{1/4} n_{10}^{-1/4}, \tag{12}$$

where $n_j = L_j/4\pi r^2 \Gamma_j^2 m_p c^3$ is the number density in the rest frame of the jet. Note that in the last expression we have assumed that $r < R_b$, in which case $\Gamma_{\rm sh}$ is independent of distance because both the density of the jet and external medium scale $\propto r^{-2}$. During this stage the jet is coasting, such that the emission from the shock is observed at a time $t_{\rm obs} \sim r/\Gamma_{\rm sh}^2 c \propto r$, beginning after the onset of the jet activity.

---

[5]Note that jet activity persists by injecting more energy at a lower rate, which may contribute to the late-time light curve.

9

In modeling the emission from the external shocks, shown in Figure S2, we adopt the gamma-ray burst afterglow model of Sari et al. (*47*). We assume that a fraction $\epsilon_e$ and $\epsilon_B$ of the jet energy is placed into that of non-thermal electrons (that follow a power-law distribution with $p = 2.5$) and the magnetic field, respectively. The time dependence of the self-absorption, characteristic, and cooling frequencies of the forward shock emission are, respectively, given by $\nu_{sa} \propto \nu_m \propto t^{-1}$, and $\nu_c \propto t$. The flux at the characteristic frequency $\nu_m$ is constant with time at the value $F_{\nu,m} = 100 \epsilon_{B,-2}^{1/2} E_{53} n_1^{1/2}$ mJy. For this choice of parameters, emission from the reverse shock is subdominant to the reverse shock because (a) the relative Lorentz factor between the the shocked fluid and unshocked jet is modest $\Gamma_{\rm rel} = 0.5(\Gamma_j/\Gamma_{sh} + \Gamma_{sh}/\Gamma_j) \simeq 1.6$; and (b) the reverse shock emission suffers substantial self-absorption.[6]

The reverse shock crosses the ejecta at a distance $R_{\rm cr} \sim \Gamma_{\rm sh}^2 \Delta \simeq 2 \times 10^{17} L_{47.5}^{1/2} n_{10}^{-1/2} \delta_{16}$ cm (*46*) similar to the assumed Bondi radius $R_b$. At radii $r > R_{\rm cr}$, the blastwave propagates into a constant density medium and reaches a self similar evolution $\Gamma_{\rm sh} \propto r^{-3/2}$. This transition occurs at an observer times that is slightly longer than the duration of the jet activity [defined as when most of the energy is ejected; (*49*)].

Once the self-similar evolution is established, the evolution of the synchrotron spectrum is described as in Sari et al. (*47*). When the synchrotron frequencies $\nu_{sa} < \nu_m < \nu_c$ they find that $\nu_{sa} = 2 \times 10^{10} \epsilon_{B,-2}^{1/5} \epsilon_{e,-1}^{-1} E_{53}^{1/5} n_1^{3/5}$ Hz, $\nu_m = 3 \times 10^{11} \epsilon_{B,-2}^{1/2} \epsilon_{e,-1}^2 E_{53}^{1/2} t_{\rm days}^{-3/2}$, and $\nu_c = 8 \times 10^{13} \epsilon_{B,-2}^{-3/2} E_{53}^{-1/2} n_1^{-1} t_{\rm days}^{-1/2}$. These expressions are valid for observer time $t \gtrsim cT_j \sim 2$ days. The flux at the characteristic frequency is $F_{\nu_m} = 170 \epsilon_{B,-2}^{1/2} E_{53} n_1^{1/2} t_{\rm days}^{-3/4}$ mJy, where we have applied a correction $(\theta_j \Gamma_{\rm sh})^2 \lesssim 1$ to the Sari et al. (*47*) expression to account for the fact the observer already observes the edges of the jet because (unlike GRB afterglows) we have assumed that initially the Lorentz factor and opening angle are related by $\theta_j = 1/\Gamma_j$.

---

[6] Note that Giannios & Metzger (*48*) assumed a much longer event duration, i.e. months not days. As a result they concluded that the reverse shock was relativistic and dominated the emission. This limit may still apply to Sw 1644+57 at late times if a lower luminosity jet persists.

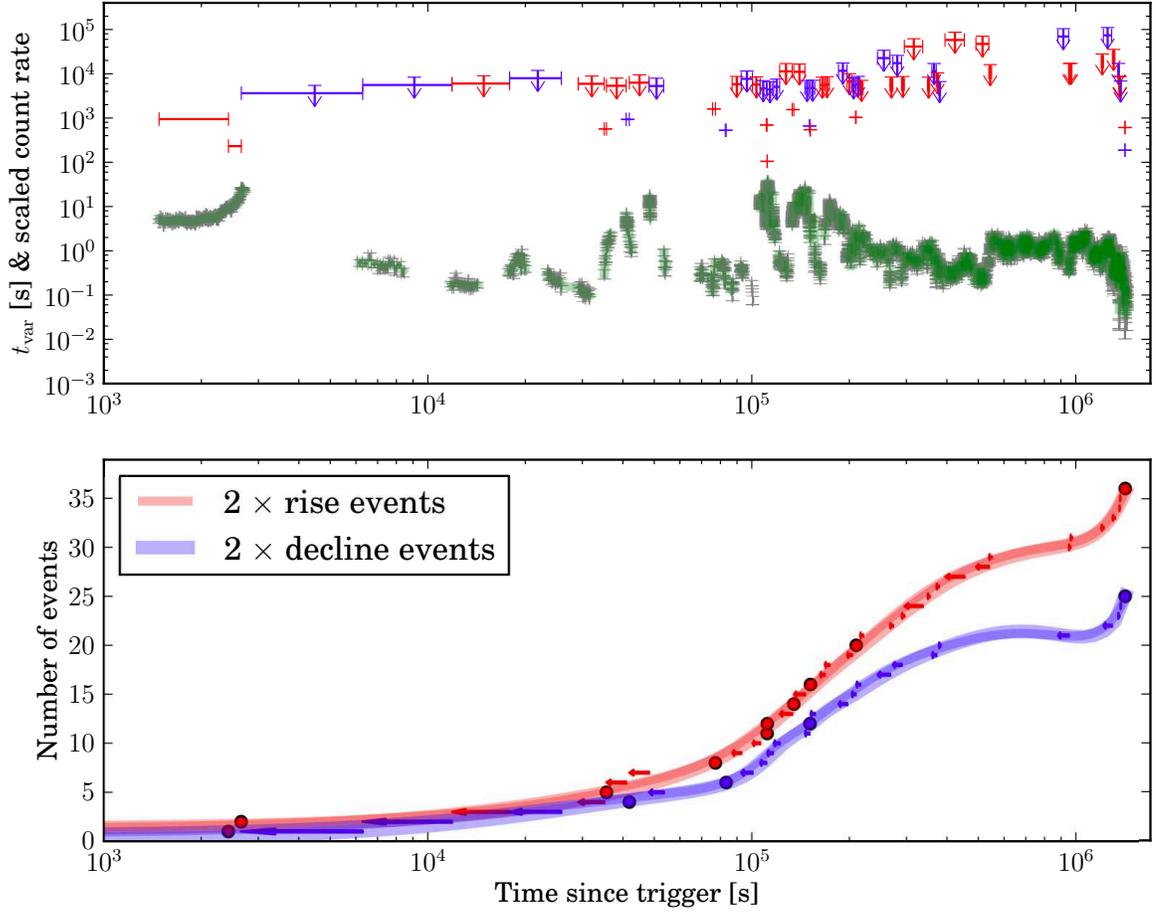

**Fig. S1a (top): Variability analysis of the X-ray light curve (gray with green spline overlay) observed with the Swift XRT. We define the variability timescale $t_{\rm var}$ as the time between count rate rising (red) or declining (blue) by a factor of 2. Upper limits on $t_{\rm var}$ can occur between gaps in the observing coverage by the satellite. (bottom) Cumulative rate of significant rising and declining events from $t_0 + 10^3$ sec to $t_0 + 1.4 \times 10^6$ sec. A number of the measured events ($t_{\rm var} \approx 100 - {\rm few} \times 10^3$) occur in the timespan from $t_0 + 3 \times 10^4$ sec to $t_0 + 3 \times 10^5$ sec.**

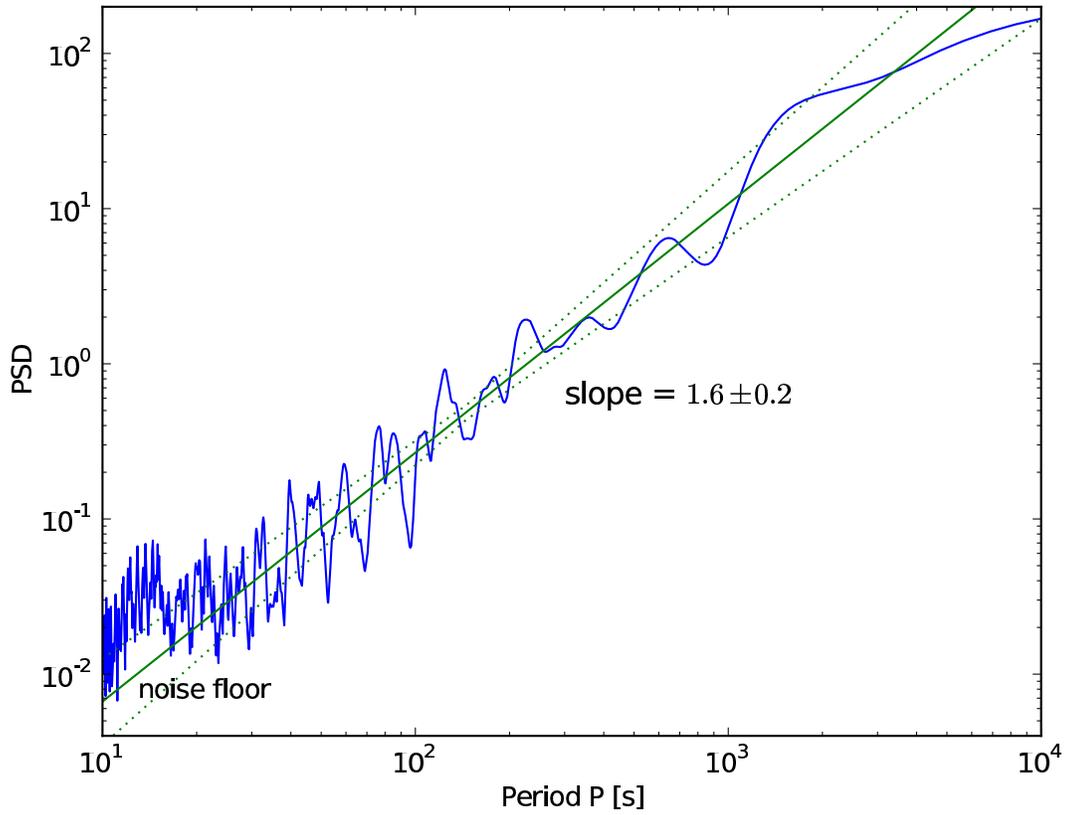

**Fig. S1b: Inferred power density spectrum of the Swift/XRT light curve. The solid straight line shows the fit to a single power ("red noise") along with the 1 $\sigma$ uncertainty range (dotted lines).**

12

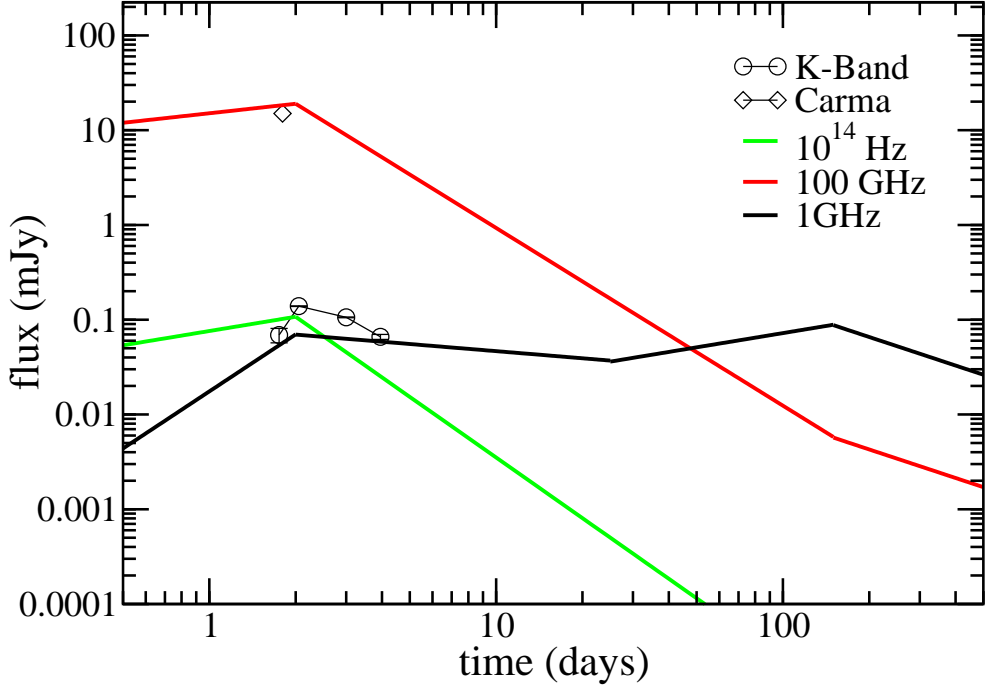

**Fig. S2: The predicted light curves from the forward shock synchrotron model at typical radio $\nu = 1$ GHz, millimeter $\nu = 100$ GHz, and IR frequencies $\nu = 10^{14}$ Hz for $E_{53} = 0.3$. The observed radio and IR light curves are overplotted. A prediction of this model is that the 100 GHz/IR emission will decline at late times. Note, however, that if the X-ray emission from J164449.3+573451 continues at a similar level, soon the energy injected by the quiescent phase will eclipse that injected in the first few days. This additional energy is not accounted for in this model and may cause the light curves shown to decline less rapidly at late times. On timescales ∼months-years the light curve will revert to the predictions of Giannios & Metzger (*48*).**

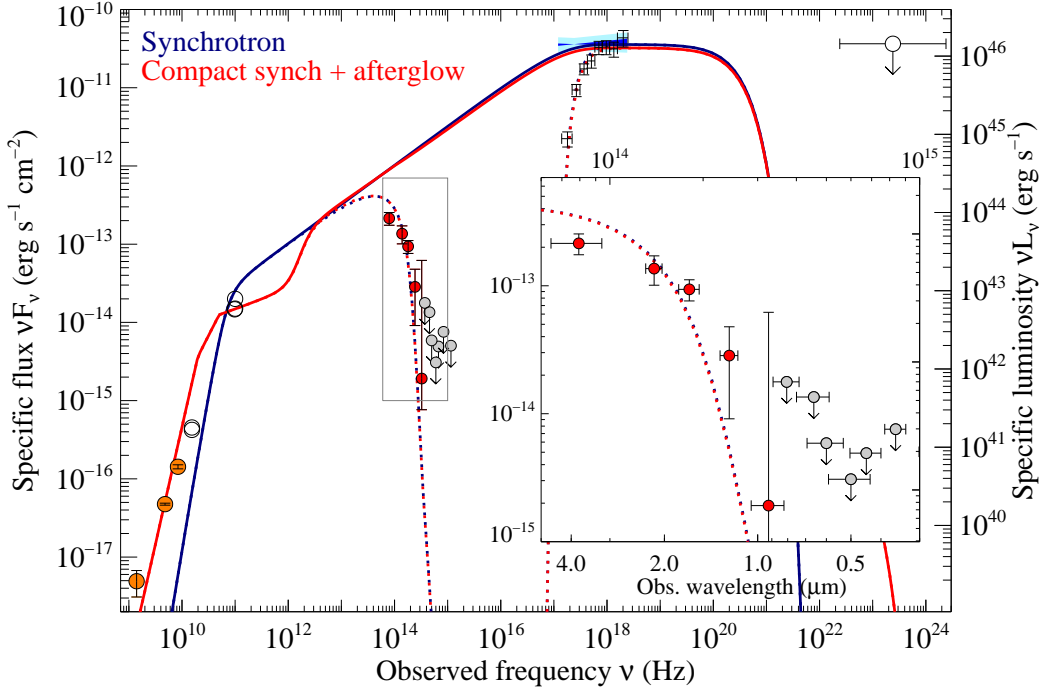

**Fig. S3: Two purely-synchrotron models fit to the instantaneous SED at $t_0 +$ 2.9 day, following from Fig 2. of the main text. The dark blue line shows a simple synchrotron spectrum matched to the data. A large extinction column ($A_V > 10$ mag) is needed to explain the relatively faint optical/NIR fluxes, and the fit is poor. In addition, the large value of the cooling frequency $\nu_c$ in this model requires a source size inconsistent with the observed variability. The red line shows a two-component synchrotron model, consisting of (1) a more compact emitting region more consistent with the optical/NIR photometry and variability timescale and (2) a more radially-extended synchrotron component, which represents synchrotron emission from the afterglow (see Section 1.8 and Fig. 1 of the main text).**



\end{document}